\documentclass[a4paper,11pt]{article}

\usepackage{times}
\usepackage{mathptmx}

\usepackage{amssymb}
\usepackage{amsthm}
\usepackage{amsmath}
\usepackage{authblk}
\usepackage{booktabs}
\usepackage	[
				figurename={Fig.},
			]{caption}
\usepackage{fancyhdr}
\usepackage	[
				paper=a4paper,
    			marginparwidth=0cm,     
    			marginparsep=0in,       
    			hmargin={1in,1in},
    			vmargin=1in,           
			]{geometry}
\usepackage[figuresright]{rotating}
\usepackage{titlesec}
\usepackage{wasysym}

\titleformat*{\section}{\bf \large}
\titleformat*{\subsection}{\itshape \normalsize}

\begin{document}

\title{A heuristic optimization method for mitigating the impact of a virus attack}

\author[a]{V.V. Kashirin\footnote{Corresponding author. Tel.: +7-965-073-0861. E-mail address: kashirin.victor@gmail.com.}} 
\author[a,b]{L.J. Dijkstra}

\affil[a]{\footnotesize National Research University of Information Technologies, Mechanics and Optics (NRU ITMO), Kronverkskiy pr. 49, 197101, Saint Petersburg, Russia}
\affil[b]{\footnotesize University of Amsterdam (UvA), Science Park 904, 1098 XH, Amsterdam, The Netherlands}

\maketitle

\begin{abstract}
\noindent Taking precautions before or during the start of a virus outbreak can heavily reduce the number of infected. The question which individuals should be immunized in order to mitigate the impact of the virus on the rest of population has received quite some attention in the literature. The dynamics of the of a virus spread through a population is often represented as information spread over a complex network. The strategies commonly proposed to determine which nodes are to be selected for immunization often involve only one centrality measure at a time, while often the topology of the network seems to suggest that a single metric is insufficient to capture the influence of a node entirely. 

In this work we present a generic method based on a genetic algorithm (GA) which does not rely explicitly on any centrality measures during its search but only exploits this type of information to narrow the search space. The fitness of an individual is defined as the estimated expected number of infections of a virus following SIR dynamics. The proposed method is evaluated on two contact networks: the Goodreau's Faux Mesa high school and the US air transportation network.
The GA method manages to outperform the most common strategies based on a single metric for the air transportation network and its performance is comparable with the best performing strategy for the high school network.  
\end{abstract}

\section{Introduction}
\label{sec:introduction}

Many viruses spread through human population by means of personal contact between infectious individuals (those who carry the virus) and susceptibles (those who are not ill at the moment but can catch the disease) \cite{Hethcote2000, Newman2002}. A concept that proved to be very valuable in order to gain a better understanding of the virus spread process is the complex network \cite{Newman2010}. Nodes within this graph structure represent individuals and are associated with a certain state (e.g., susceptible or infectious). Edges between these nodes account for social interactions. The percolation of the virus through the population (network) is then predicated by a fixed set of rules \cite{Newman2002, Boccaletti2006, Pastor-Satorras2001, Tao2006}. 

This approach has not only been employed in order to better understand the dynamics of such a disease spread, but also gives rise to a field of research with a more proactive attitude: which individuals in the population should be immunized to limit the spread of the virus most effectively? Or, to frame it a bit differently, which nodes in the network should be protected in order to limit the damage done as much as possible? 

\begin{figure}[t!]
	\centering
	\includegraphics[width=.6\textwidth]{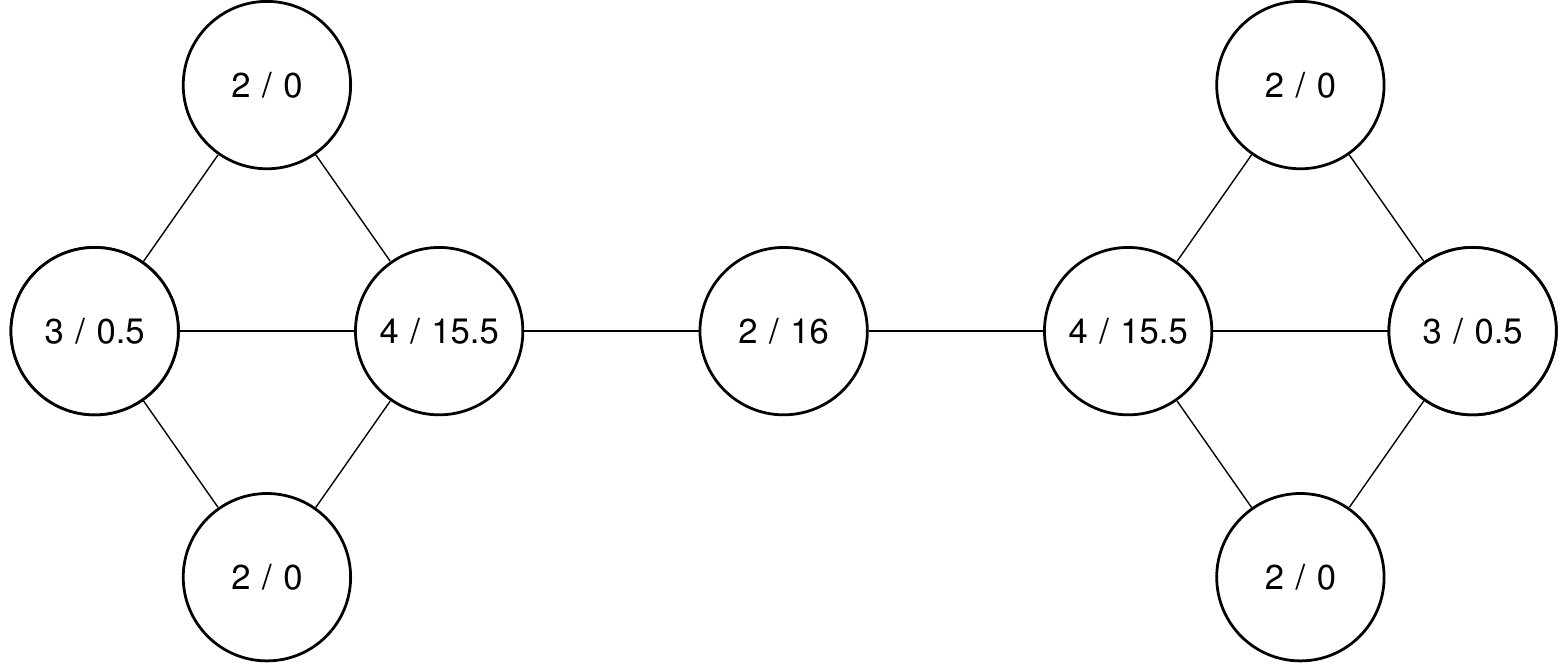}
	\caption{A rather simple hypothetical network. The values in each node represent the node's degree and betweenness centrality, respectively.}
	\label{fig:example}
\end{figure}

The concept commonly used in order to find those nodes that are in need of protection (or removed in some cases) is \emph{node centrality}\footnote{Measures for edge centrality exist as well, see \cite{Newman2010}.} \cite{Holme2002, Memon2006, Kitsak2010, Bright2011, Hou2012, Chen2012}. Node centralities express to what extent the node facilitates the spread of information over the network (note that information spread is, to some extent, alike to a virus spread \cite{Daley1964}). In this article we consider three common traditional variants\footnote{More complicated centrality measures have been proposed as well, see, for example, \cite{Kitsak2010, Hou2012, Chen2012}.} used in the literature, each of which formalizes the concept of centrality in a (slightly) different manner: 1) \emph{degree centrality} which is equal to the node's degree, 2) \emph{betweenness centrality} which focusses rather on to what extent the node could influence the communication between other nodes in the network, and 3) \emph{eigenvector centrality} which expresses the relative importance of a node in terms of importance of its direct neighbors (a having a few connections with important nodes is more important than the one having the same number of connections with less important ones). The definitions of these centrality measures are given in section \ref{sec:centrality}. The usual approach to locate those nodes that influence the spread of information the most is to determine for all nodes in the network their score on one centrality measure only. A fraction of nodes that scored the highest are then proposed to be immunized. This approach has its drawbacks. Take, for example, the network in Fig.~\ref{fig:example}. It consists of two clusters connected by one single node. If one would only use a ranking based on degree centralities (as it rather common in the literature), one would start by immunizing several nodes in the clusters. The node in the center of the graph is left untouched, while immunizing it at an early phase, would prohibit the virus to spread from one cluster to the other. The importance of this node would be noted only when one would take betweenness centralities into account as well. Of course, this example is rather simple and one would be able to come up with such an analysis by simply examining the graph visually. When the network gets large, however, (e.g., 500 nodes as in one of the networks examined later -- see section \ref{sec:networks}) it is rather hard to make these kinds of assertions and a more structural approach is required. 

In this paper we propose a genetic algorithm \cite{SivanandamS.N.andDeepa2007} (GA) for finding those individuals that need to be protected in order to limit the spread of the virus through the rest of the network as much as possible. This heuristic search approach differs from the methods proposed in the literature, since the GA is allowed to search through a wide range of subsets of nodes and does not rely explicitly on any kind of centrality metrics during its search; centrality measures are only used to narrow the search space. Nodes that score low on all measures of centrality most likely do not play an important role in the spread of the virus and can, thus, be neglected.  

In order to emulate the spread of the virus over the network we define an adapted version of one of the most common models in epidemiology: the \emph{SIR model} \cite{Hethcote2000, Newman2002, Newman2010, Daley1964}. The model was first introduced in the twenties by Lowell Reed and Wade Hampton Frost. They proposed to divide the entire population into the following three distinct classes:  
\begin{itemize}
	\item The \emph{susceptibles}, $S$: the group of individuals who have not been infected but can catch the disease.
	\item The \emph{infectives}, $I$: the group consisting of all individuals that are currently infected by the disease and could infect others from the susceptible class. 
	\item The \emph{removed}, $R$ (sometimes referred to in the literature as recovered): those who had the virus but either recovered and gained immunity or died.  
\end{itemize}
Note that an individual in the model can go through three strictly sequential phases: susceptible individuals can get infected by the virus and either recover or die; removed individuals never loose their gained immunity\footnote{Several extensions of the models exist, e.g., the SIRS model where recovered individuals can return to the susceptible state. See the paper by H.W. Hethcote (2000)\nocite{Hethcote2000} for a splendid overview of these so-called compartmental models.}. In section \ref{sec:sir} we discuss our adaptation of this model for simulating the spread on networks. 
  
The structure of the paper is organized as follows. In the following section we discuss the methods used. First we define the optimization task at hand formally in section \ref{sec:formal}. Section \ref{sec:networks} introduces the two networks that are used for validation of the methods. We proceed with defining the three different centrality measures discussed earlier in the text. In section \ref{sec:sir} we formally define an adaption of the traditional SIR compartmental model for network structures. Section \ref{sec:ga} contains a detailed description of the genetic algorithm for finding the optimal set of nodes that need to be protected using the various centrality rankings. The results are presented in section \ref{sec:results}. We finish this paper with our conclusions, some discussion and a few pointers for future research. 

\section{Methods}
\label{sec:methods}

\subsection{The optimization task}
\label{sec:formal}

The optimization task presented in the introduction can be expressed a bit more formally in the following manner. Suppose we are dealing with a network $G = \left<V, E\right>$ where $V$ denotes the set of vertices in the network and $E$ is the set of (bidirectional) edges between pairs of nodes in $V$. The size of the network is given by $N = |V|$. The number of nodes that can be protected is limited due to time and resource constraints to a total of $k$ nodes. We are, thus, interested in finding a $k$-subset of nodes $I \subseteq V$ that, when protected, will most effectively limit the spread of the virus through the network $G$. This task is by no means easy. The number of possible $k$-subsets of nodes is $\binom{N}{k}$ which, for most real-world networks, is simply too large to brute-force. In addition, one is generally unaware where the virus starts which needs to be accounted for when assessing the successfulness of a solution. 

In the following sections we describe a method to find a (near-)optimal $k$-subset of nodes. 

\subsection{The data sets used for validation}
\label{sec:networks}

In this paper we consider two contact networks to validate our approach, one social and one transportation network:  
\begin{enumerate}
	\item The Goodreau's Faux Mesa high school network which is the result of a simulation of a in-school friendship network based on data from a high school in the rural western United States \cite{Resnick1997}. The network consists of $147$ nodes and $202$ undirected edges (representing mutual friendship).  See Fig.~\ref{fig:topo}a for a visual representation of the graph. 
	\item The US air transportation network \cite{Colizza2007}, which consists of $500$ nodes (US airports with the highest amount of traffic) and $2980$ undirected edges (representing air travel connections). See Fig.~\ref{fig:topo}b for a visual representation of the graph. 
\end{enumerate}
Both data sets are freely available and can be found, respectively, at the website of the CASOS group\footnote{http://www.casos.cs.cmu.edu/computational\_tools/datasets/external/Goodreau/index11.php} and the Cx-Nets website\footnote{http://sites.google.com/site/cxnets/usairtransportationnetwork}.

Please, note that the models, methods and measures discussed in the following sections can be applied (or extended) to any network structure. The networks chosen here are only used as examples. 

\subsection{Node centrality}
\label{sec:centrality}

Measures for centrality try to formalize the (relative) importance of a node in the network. In this paper we consider the following usual measures: 1) degree, 2) betweenness and 3) eigenvector centrality. 

Degree centrality, $C_D$, formalizes the importance of a node by setting it equal to the number of neighbors the node has; the more connections, the more important the node:  
\begin{equation}
	C_D (v) = \mathrm{deg}(v). 
	\label{eq:d}
\end{equation}
Betweenness centrality, $C_B$, tries to capture a different kind of `importance'. The idea behind this formalism is that the importance of a node depends on to what extent it can influence the communication between the other nodes in the network \cite{Newman2010}. The measure is computed by determining the fraction of the shortest paths between each pair of nodes in the network that contains the node $v$, i.e.: 
\begin{equation}
	C_B (v) = \displaystyle\sum_{s\neq v\neq t \in V} \frac{\sigma_{s,t} (v)}{\sigma_{s,t}},
	\label{eq:b}
\end{equation}
where $\sigma_{s,t}$ is the number of the shortest paths between the nodes $s$ and $t$ and $\sigma_{s,t}(v)$ returns the number of shortest paths that passes through the vertex of interest, $v$. As one can imagine, this measure for centrality is computationally rather hard to determine. In 2001\nocite{Brandes2001} Ulrik Brandes presented a fast version of the algorithm; computation time is upper bounded by $O(|V||E|)$. We will use this algorithm for determining the betweenness centrality throughout this paper. 

Eigenvector centrality, $C_E$, distinguishes itself from the other measures discussed here, since it takes the importance of its neighbors explicitly into account:
\begin{equation}
	C_E (v) = \frac{1}{\lambda} \displaystyle\sum_{w \in V} a_{v,w} C_E (w),
	\label{eq:e1}
\end{equation}
where $a_{v,w}$ is the $(v,w)$-entry of the adjacency matrix of the network in question. Rewriting this expression to matrix form yields
\begin{equation}
\mathbf{A}\mathbf{x} = \lambda \mathbf{x},
	\label{eq:e2}
\end{equation} 
where the $i$-th entry of  $\mathbf{x}$ is equal to $C_E(i)$. In other words, the eigenvector centrality of a node $v$ is the $v$-th entry of the  right eigenvector associated with the first eigenvalue $\lambda$ of the adjacency matrix $\mathbf{A}$ of the network. 

\subsection{An adaption of the SIR model for networks}
\label{sec:sir}

The SIR model is one of the most often used models for simulating the spread of a virus through a population. 
The original model in terms of a set of ordinary differential equations makes the assumptions that 1) the population size, $N$, is large and 2) the population is `fully mixed', meaning that every individual has the same number of (randomly picked) connections and have (approximately) the same number of contacts at the same time \cite{Hethcote2000, Newman2002}. Of course, these assumptions are rather unsatisfactory and can be overcome by extending the SIR model to networks \cite{Newman2002, Newman2010}.  

Suppose we have the following network $G = \left<V,E\right>$ where the set $V$ denotes the vertices of the network and $E$ is the set of bidirectional edges between pairs of vertices in $V$. Each node $v$ in the set of vertices $V$ is associated with either the state $S$ (susceptible), $I$ (infectious) or $R$ (removed) at each moment in time, i.e., $\mathrm{St}\left( v, t\right) \in \left\{S, I, R\right\}$. The initial state of the network is given by $\{S_0 , I_0 , R_0 \}$, i.e., the sets of nodes who's state at the start of the simulation, $t = 0$, are either $S, I$ or $R$. The dynamics of the virus are emulated by updating the state of all the nodes in the network simultaneously while stepping forward in time. 
When the state of node $v \in V$ at moment $t$ is susceptible, i.e., $\mathrm{St}(v, t) = S$, the state of this node at the next time step is given by  
\begin{equation}
\mathrm{St}\left(v, t + 1 \right) := \begin{cases}
		I & \text{with probability }1 - (1 - \beta)^{| \mathcal{N}_v(I) |}, \\ 
		S & \text{otherwise,}
	\end{cases} 
	\label{eq:s}
\end{equation}
where $\beta$ is the chance of the susceptible node to be infected by a single infectious neighbor. The function $\mathcal{N}_v(I)$ returns the set of infectious neighbors in the direct vicinity of node $v$. The operator $|.|$ returns the cardinality of a set. Note that the chance of \emph{not} becoming infected decreases exponentially with the number of infectious neighbors. 
 
When node $v$ is at time $t$ infective, its state becomes
\begin{equation}
\mathrm{St}\left(v, t + 1\right) := \begin{cases}
		R & \text{with probability }\gamma, \\ 
		I & \text{otherwise,}
	\end{cases} 
	\label{eq:i}
\end{equation}
where $\gamma$, thus, denotes the chance of recovering (or dying) from the virus during one time step. 
Since we assumed that once a node reached the removed state $R$, it is either indefinitely immune for the virus or dead, the state of that node will not change till the end of the simulation. The simulation ends when there are no more infective nodes in the networks; the spread of the virus grinded to a halt and the number of susceptible and removed nodes in the network will not change. The total number of casualties is equal to the number of nodes in state $R$. 

\subsection{Genetic algorithm} 
\label{sec:ga}

In order to find a (near-optimal) set of nodes that need to be protected from the virus we employ a genetic algorithm \cite{SivanandamS.N.andDeepa2007}. First, we compute the degree, betweenness and eigenvector centrality for each node in the network (see section \ref{sec:centrality}). By ranking the nodes according to their centrality scores, we obtain the following three ascending rankings: 
\begin{equation}
	R_D = \{v_{(1)}^D, v_{(2)}^D, \dots, v_{(N)}^D\}, \quad R_B = \{v_{(1)}^B, v_{(2)}^B, \dots, v_{(N)}^B\} \quad \text{and} \quad R_E = \{v_{(1)}^E, v_{(2)}^E, \dots, v_{(N)}^E\}, 
	\label{eq:rankings}
\end{equation}
where $R_D$, $R_B$ and $R_E$ are the rankings based on, respectively, the degree (\ref{eq:d}), betweenness (\ref{eq:b}) and eigenvector (\ref{eq:e1}) centrality. We limit the search space of the GA by taking into account that nodes with low centrality scores on all three centrality measure do most likely not play an important role in spreading the virus over the network. We, thus, keep only the first $l < N$ nodes in each ranking, i.e., 
\begin{equation}
	R'_D = \{v_{(1)}^D, v_{(2)}^D, \dots, v_{(l)}^D\}, \quad R'_B = \{v_{(1)}^B, v_{(2)}^B, \dots, v_{(l)}^B\} \quad \text{and} \quad R'_E = \{v_{(1)}^E, v_{(2)}^E, \dots, v_{(l)}^E\}. 
	\label{eq:r}
\end{equation}
The genetic algorithm will only search for an optimal $k$-subset of nodes in the union of these reduced rankings:
\begin{equation}
	R' = R'_D \cup R'_B \cup R'_E. 
	\label{eq:reduced}
\end{equation}
Each individual in the GA is represented by a chromosome with $k$ genes:
\begin{equation}
	I = \{v_1, v_2, \ldots, v_k\} \qquad \text{where }v_i \in R'.
	\label{eq:individual}
\end{equation}
The genes, thus, represent the nodes that need to be protected from the virus where we only take into account those nodes that score relatively high on at least one of the centrality measures. 

In order to determine the fitness of an individual, firstly we take the original network $G = \left<V, E\right>$ and remove the nodes present in the chromosome $I$. (Removing nodes corresponds here to immunizing the nodes from the virus). Secondly, in order to account for the fact that one is normally unaware where the virus starts, we select one node at random in the new graph to be infectious while keeping the other nodes susceptible. Thirdly, we apply the SIR model as described in section \ref{sec:sir} and store the number of removed individuals at the end of the simulation which we denote with $N_{\text{casualties}}$. (Recall that the simulation ends when there are no more infectious nodes in the network). Since the selection of infectious individuals is a stochastic process we repeat the last two steps $m$ times. The fitness of an individual in the GA population is then defined as the expected number of casualties: 
\begin{equation}
	\text{fitness}(I) = \mathrm{E}\left[N_{\text{casualties}} \right] \approx \frac{1}{m} \displaystyle\sum_{i = 1}^{m} N_{\text{casualties}}^{(i)}
	\label{eq:fitness}
\end{equation}
where $N_\text{casualties}^{(i)}$ is the number of recovered nodes at the end of the $i$-th simulation of the SIR model applied to the updated graph $G$. 

\begin{table}[t!]
	\centering
	\caption{Parameter settings.}
	\scriptsize{
	\begin{tabular}{l l c}
		\toprule
		Symbol & Description & Value range\\
		\midrule
		$\beta$ & The chance of becoming infected by one infectious neighbor during one time step, see eq. (\ref{eq:s}) & .3\\
		$\gamma$ & The chance of recovering during one time step, see eq. (\ref{eq:i}) & .3 \\
		\midrule
		$k$ & The number of nodes that are to be protected from the virus & $\{10, 50\}$ \\
		$l$ & The number of nodes with the highest scores on a centrality measure that should be kept, see section \ref{sec:ga} & $\{100, 200\}$\\
		$m$ & The number of SIR simulation runs for the determining the fitness of an individual in the GA, see eq. (\ref{eq:fitness}) & 100 \\ 
		\midrule
		- & The size of the GA population & $100$ \\
		- & The number of generations simulated by the GA & $100$ \\
		- & The size of the tournament used during tournament selection & $4$ \\
		- & The mutation rate for a gene & $1/k$ \\   
		\bottomrule
	\end{tabular}
	}
	\label{tab:parameters}
\end{table}

In summary, the objective of the GA is find that subset $I^\ast$ that minimizes the expected number of casualties\footnote{Although it is more common to maximize a fitness function rather than minimizing it, we felt minimization would be more appropriate, since we are dealing with the expected number of casualties.}: 
\begin{equation}
	I^\ast = \operatorname*{arg\,min}_{I \subseteq R', |I| = k} \text{fitness}(I).
\end{equation}

The genetic algorithm simulates $100$ generations each with a total of $100$ individuals. Selection for mating is performed by applying tournament selection where the tournament size is set to $4$. As crossover operator we use an adaptation of uniform crossover. Suppose we selected two parents to mate, e.g., $P_1 = \{1,2,3,4\}$ and $P_2 = \{3,4,5,6\}$. We then concatenate the chromosomes of these parents and sort the resulting array. In our example, we find $\{1,2,3,3,4,4,5,6\}$. The chromosome of the first child, $C_1$, consists of the odd entries of the array; the chromosome of the second child, $C_2$, consists of the even entries, e.g., $C_1 = \{1,3,4,5\}$ and $C_2 = \{2,3,4,6\}$. This approach guarantees the absence of duplicates in the new chromosomes. The mutation rate per gene is set to $1/k$, i.e., each chromosome undergoes on average one mutation in one gene every generation. The mutation of a gene entails that its value is randomly set to a node in the set $R'$ that is not in the individual's chromosome already.

The initial population is randomly generated except for three individuals: their chromosomes are set to the first $k$ nodes in, respectively, the degree, betweenness and eigenvector centrality rankings as given in eq. (\ref{eq:r}). In addition, we keep for every generation the $10$ best performing individuals from the previous generation in order not to loose good solutions that were found earlier. 

\section{Results}
\label{sec:results}

Table \ref{tab:parameters} provides an overview of the parameters used in the SIR model and the GA introduced earlier. In addition, it presents the parameter settings that were used for producing the results presented in this section. 

\begin{figure}[t!]
	\centering
	\includegraphics[width = \textwidth]{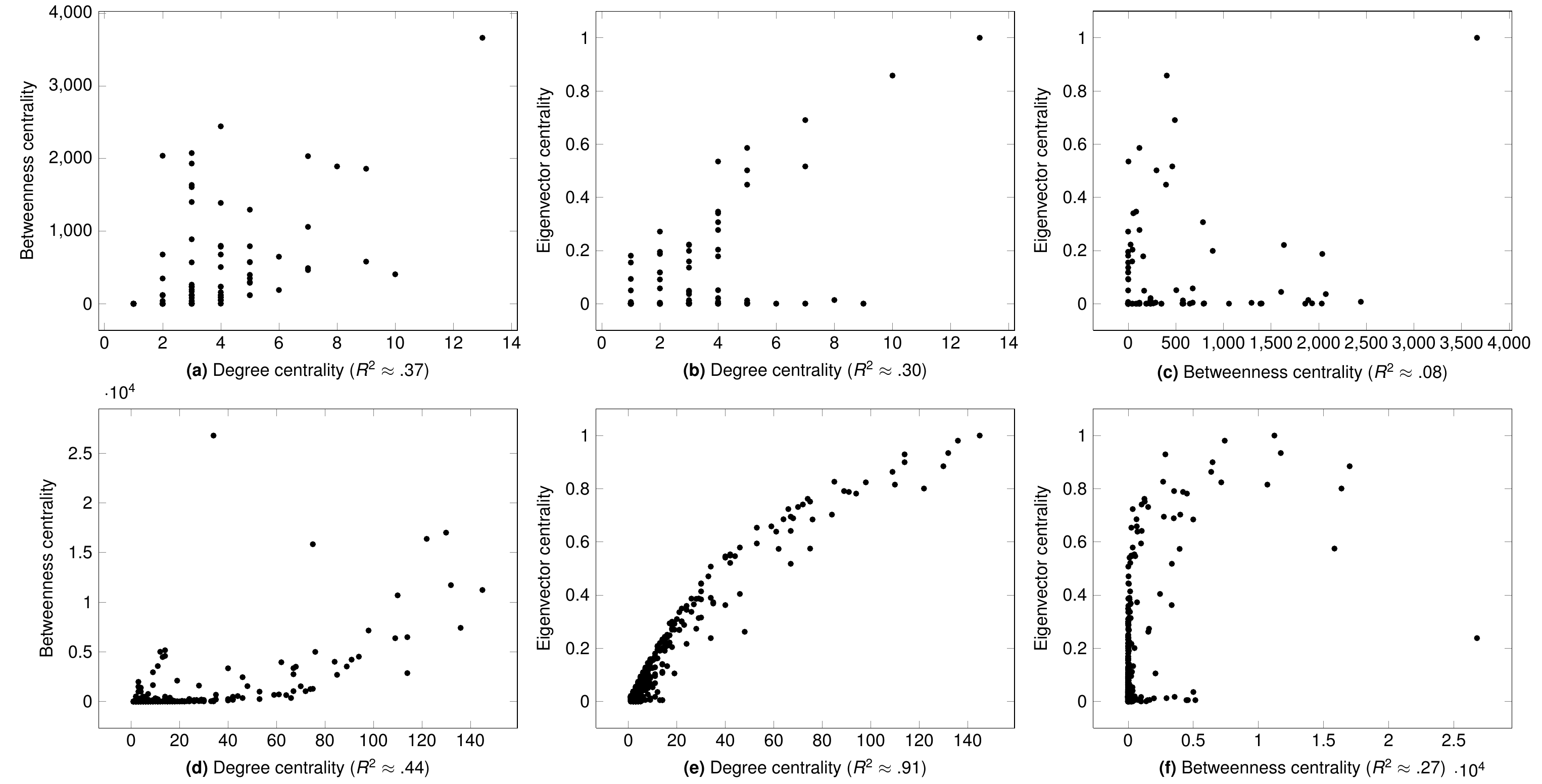}
	\caption{The correlations between the degree, betweenness and eigenvector centralities in both networks. The scatter plots (a)-(c) represent the data from the Goodreau's Faux Mesa high school network, and the scatter plots (d)-(f) depict the data from the air transportation network.}
	\label{fig:corr}
\end{figure}

Fig.~\ref{fig:corr} depicts the correlations between the degree, betweenness and eigenvector centralities for the nodes in both the networks. The reported $R^{2}$'s are determined by applying linear regression. Note that in both cases degree centrality seems to correlate with its betweenness and eigenvector counterparts while betweenness and eigenvector centralities do not seem to correspond. It is interesting to see that degree and eigenvector centralities seem to correlate more for the air transportation network (e) than for the high school network (b). This might be due to the higher number of edges in the air transportation network (2980 against 500 nodes) in contrast to the high school network (202 edges against 147 nodes). The main point of these figures is to show that there are discrepancies between the various centralities measures, which the genetic algorithm from section \ref{sec:ga} is set out to exploit in order to find a more optimal set of nodes. 

\begin{figure}[b!]
	\centering
	\includegraphics[width = \textwidth]{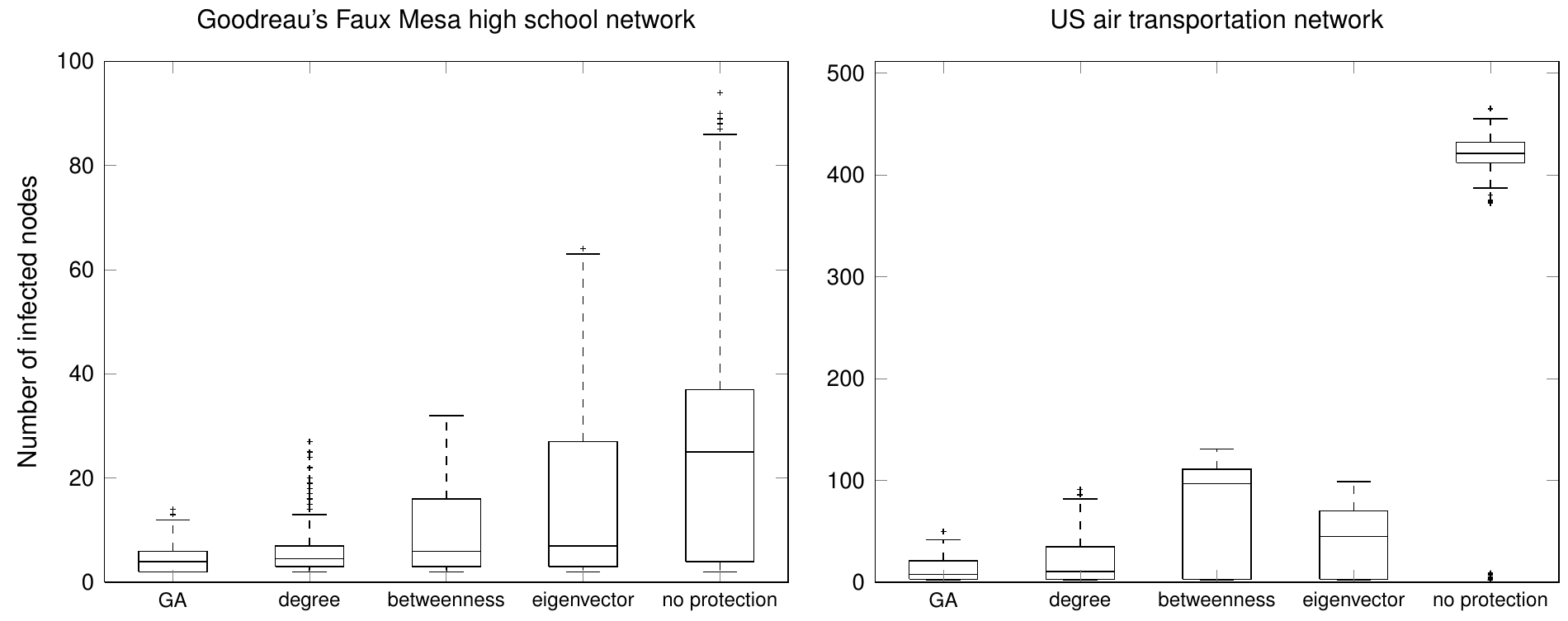}
	\caption{The boxplots for the number of infected nodes found when various strategies are applied. GA stands for the genetic algorithm. `degree', `betweenness' and `eigenvector' stands for the cases when only the nodes with the highest score on that particular centrality measure where protected from the virus. `no protection' refers to the case when no preparations were undertaken. The whiskers represent the lowest datum still within 1.5 IQR of the lower quartile and the highest datum still within 1.5 IQR of the upper quartile. (IQR stands for interquartile range). Outliers are denoted by a cross.}
	\label{fig:boxplot}
\end{figure}

So, to what extent does combining centralities measures help in finding a more optimal set of nodes that need to be protected from the virus? In order to answer this question, we compare four different strategies: 
\begin{enumerate}
	\item Provide protection for the nodes with the highest degree centrality, see eq. (\ref{eq:d}); 
	\item Provide protection for the nodes with the highest betweenness centrality, see eq. (\ref{eq:b}); 
	\item Provide protection for the nodes with the highest eigenvector centrality, see eq. (\ref{eq:e1}); 
	\item Provide protection for the nodes found be the genetic algorithm, see section \ref{sec:ga}. 
\end{enumerate}
In order to keep the comparison fair, each strategy is allowed to protect exactly $k$ nodes. The parameters $k$ and $l$ are set to $10$ and $100$, respectively, for the high school network and to $50$ and $200$ for the air transportation network (see Table \ref{tab:parameters}). 

\begin{figure}[t!]
	\includegraphics[width=\textwidth]{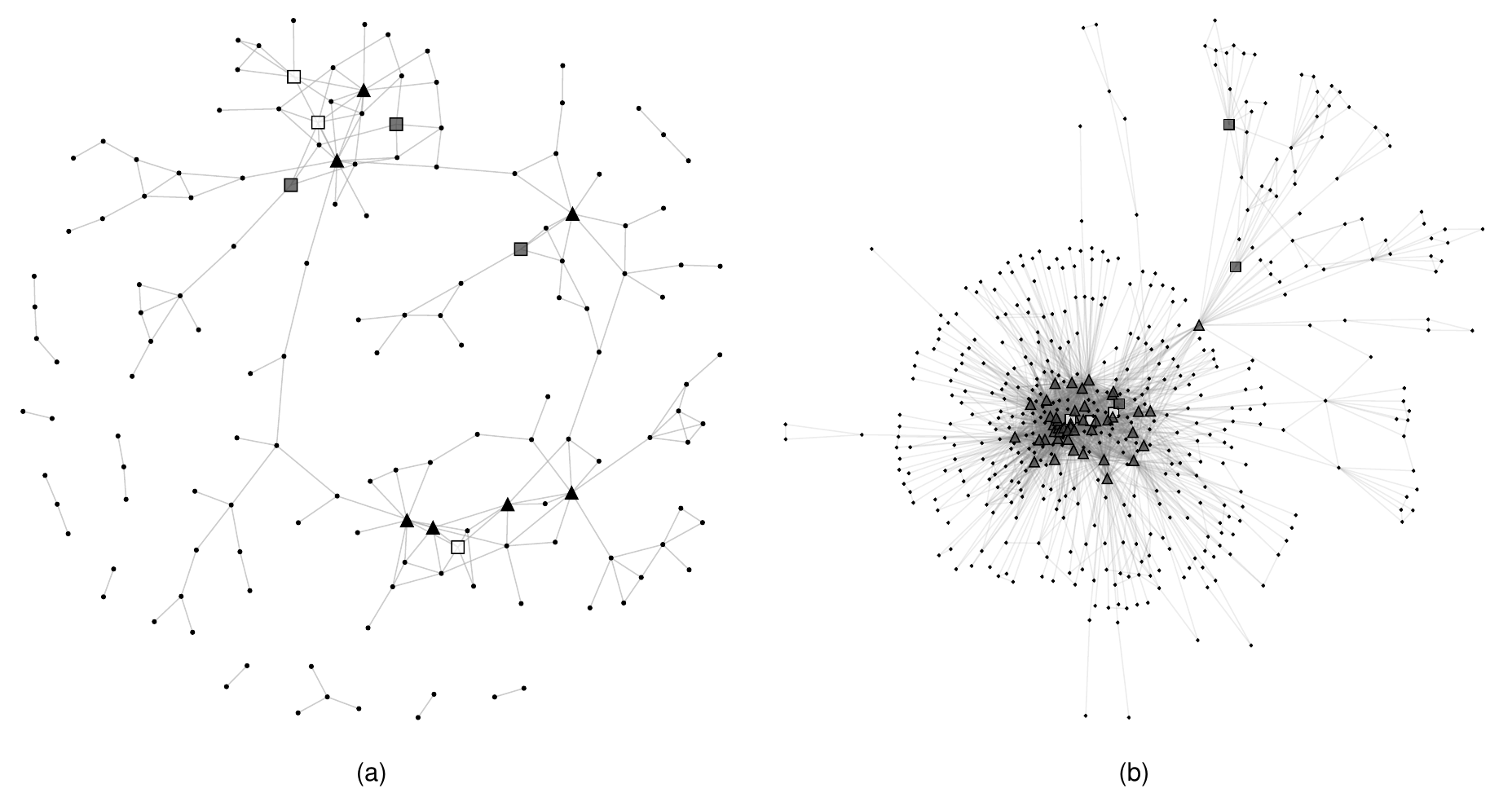}
	\caption{The topology of the Goodreau's Faux Mesa high school network (a) and the US air transportation network (b). The nodes depicted with \UParrow\ are the nodes that were suggested by both the genetic algorithm and the strategy based on degree centrality. The nodes denoted with $\blacksquare$ were only selected by the GA and not the degree strategy. Symbol $\square$ denotes the nodes that were only selected by the degree strategy. }
	\label{fig:topo}
\end{figure}

\begin{figure}[b!]
	\centering
	\includegraphics[width=.9\textwidth]{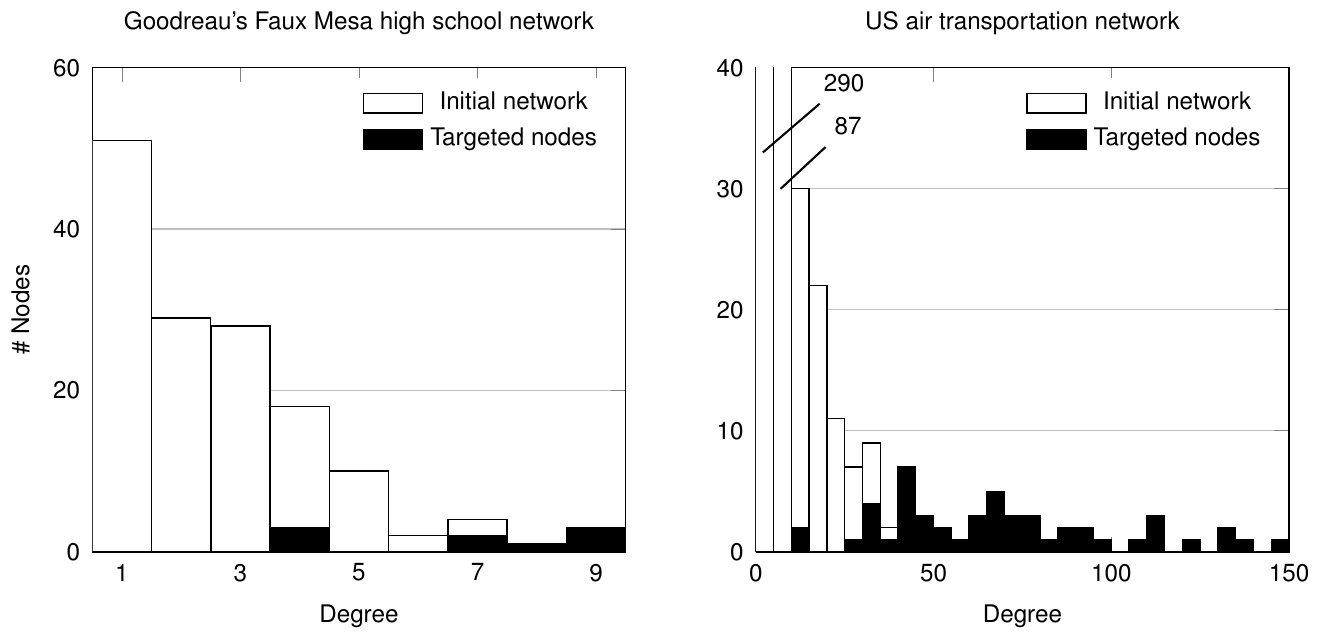}
	\caption{The degree distributions of the high school and air transportation networks. The bars marked black denote the number of nodes with that particular degree that were targeted by the genetic algorithm.}
	\label{fig:degreeDistr}
\end{figure}

After applying a strategy on a network, we run $500$ SIR simulations and count the number of nodes that got infected by the virus. Before each SIR run we randomly select a single node to be infectious (the others are susceptible) in order to account for the fact that one is normally unaware where the virus starts.
The results are presented in Fig.~\ref{fig:boxplot}. The `no protection' case represents the situation when no preparations were undertaken. For both networks this would have disastrous consequences. In the air transportation network, almost all nodes will definitely get infected. The genetic algorithm and the degree centrality approach seem to be preferred for both networks. The GA seems to outperform the degree strategy, but only slightly. The methods based on betweenness and eigenvector centralities do improve the situation, but are clearly not optimal. Note that the performance of these two approaches switch places between the graphs. The reason why eigenvector centralities might outperform the betweenness approach when applied to the air transportation network might lie in the fact that degree and eigenvector centrality expose a higher correlation in this graph than in the high school network, see Fig.~\ref{fig:corr}b and \ref{fig:corr}e.  

The genetic algorithm outperforms the strategy based on degree centrality only in the case of the US air transportation network. For the high school network their performances are comparable. Fig.~\ref{fig:topo} presents the topology of both networks and the differences between the sets of nodes proposed by the GA and the degree centrality strategy. The nodes depicted with \UParrow\ are the ones that were suggested by both strategies. The nodes depicted with $\blacksquare$ were only selected by the GA and not the degree strategy. Symbol $\square$ denotes the nodes that were only selected by the degree strategy. Note that the GA manages to locate nodes that when protected will help to `separate' clusters, i.e., it becomes harder for the virus to move from one cluster to the next. 

Fig.~\ref{fig:degreeDistr} depicts the degree distributions of both the high school and the air transportation network. The bars marked black denote the number of nodes with that particular degree that were targeted by the genetic algorithm. Note that the nodes selected by the GA are not all in the tail but also appear at the beginning and in the middle of the degree distribution. 

\section{Conclusions \& Discussion}
\label{sec:conclusions}

The work presented in this paper was undertaken to design and evaluate the effectiveness of a heuristic optimization method for virus spread inhibition, which, in contrast to earlier research \cite{Holme2002, Memon2006, Bright2011, Hou2012, Chen2012, Kitsak2010}, does not rely explicitly on any centrality measures during its search. 

We found that the genetic algorithm proposed in this paper seems to outperform the standard approaches in the literature. This is a remarkable feat since the genetic algorithm searches through the immense set of possible subsets of nodes and does not base its decision directly on various centrality rankings. Fig.~\ref{fig:topo} shows that the genetic algorithm is able to find an adequate subset of nodes that need to be protected: it is able to construct a set of nodes that not only scores high on degree, but also on betweenness centrality. The virus is, thus, restricted in its movements within a cluster (degree) but also from one cluster to the other (betweenness). 

The degree distributions in Fig.~\ref{fig:degreeDistr} show the interesting result that the nodes with the biggest influence on the spread of the virus over the network are not necessarily to be found in the tail of the degree distribution; among them there are the nodes with no high degree at all, still, by being immunized, they seem to reduce the number of infected significantly. Future research is required to identify why exactly these nodes are of importance, and, most importantly, whether there are ways to simplify the identification of these individuals. Perhaps (non-topological) node attributes can provide important clues for the selection of individuals to be immunized? Being able to locate these individuals effectively without the need for the full topology of the network will aid tremendously in mitigating the impact of the virus on a population. 

The developed method could be applied rather easily to a wide variety of situations that require urgent computing since the method is not limited to a specific type of network: (near-)optimal solutions can be found for both regular and heterogenous networks. In addition, one does not have to restrict to one particular centrality measure. The search space of the genetic algorithm can be seamlessly extended for various fields of applications, regardless of the type of metrics used. 

The presented method is recommended for topologically heterogenous networks that are  characterized by the absence of strong correlations between the various node centralities and the presence of modularity. The authors expect that for homogenous networks the traditional degree centrality approach (i.e., immunizing the nodes with the highest degree) is to be preferred. In addition, the GA requires more computation time. Especially when time is pressing, one is, therefore, recommended to resort to the traditional and fast-to-compute strategies. 

Further research might be done to explore different sets of metrics and heuristics in order to widen or narrow the search space. In addition, it might be interesting to take edge centralities into account as well, although the authors expect that this would not make a significant difference, since edge and node centralities are often highly correlated. 
The parameters for the genetic algorithm were set rather intuitively. Better results might be obtained when different parameter settings are applied.  
In this paper we only explored one particular compartmental model, the SIR model (see section \ref{sec:sir}). The method can be extended to different kind of models as well \cite{Hethcote2000, Newman2010}. It would be interesting to see how the method proposed here would perform when different virus dynamics are taking into account. The applicability and performance of other optimization methods such as simulated annealing might be a subject of future research as well. 

This paper has shown the promise of applying a genetic algorithm in order to mitigate the impact of a virus attack on a population. The authors would like to express their hope that continuing research along the lines set out in this paper will assist in gaining a better understanding of virus inhibition, and will ultimately provide better ways to avert future epidemics. 
 
\section*{Acknowledgements}
The authors thank Prof. dr. A.V. Boukhanovsky from NRU ITMO for his insightful comments and support.  
This work is supported by the \emph{Leading Scientist Program} of the Russian Federation, contract 11.G34.31.0019.

\footnotesize{
\bibliographystyle{elsarticle-num}
\bibliography{ArXiv_KashirinDijkstra}
}
\end{document}